\documentclass[12pt]{article}
\input epsf
\newcommand{\sect}[1]{\section{#1}\setcounter{equation}{0}}

\begin{document}
\date{8th August, 2000}
\def\be{\begin{equation}}
\def\ee{\end{equation}}
\def\bea{\begin{eqnarray}}
\def\eea{\end{eqnarray}}
\def\Im{{\rm Im }}
\def\Re{{\rm Re }}
\def\N{{\cal N}}
\def\F{{\cal F}}
\def\tQ{{\tilde Q}}
\def\tr{{\rm tr}}
\def\cF{{\cal F}}
\def\cL{{\cal L}}
\def\SU{{\rm SU}}
\def\U{{\rm U}}
\def\bZ{{\mathbb{Z}}}
\def\V{{\cal V}}
\def\tB{{\tilde{B}}}
\def\ie{{\it i.e.,}}
\def\ls{{\ell_s}}
\def\tF{{\tilde{F}}}
\def\tC{{\tilde{C}}}
\def\bP{{\mathbb{P}}}
\def\gs{g_{\rm s}}
\def\bs{\vspace{5pt}}
\def\ba{\begin{array}}
\def\ea{\end{array}}
\def\half{\mbox{\scriptsize{${{1}\over{2}}$}}}
\def\eighth{\mbox{\scriptsize{${{1}\over{8}}$}}}
\def\arccosh{\mbox{arccosh}}
\renewcommand\bZ{{\bf Z}}
\renewcommand\bP{{\bf P}}
\newcommand\bR{{\bf R}}
\newcommand{\OL}[1]{ \hspace{1pt}\overline{\hspace{-1pt}#1
   \hspace{-1pt}}\hspace{1pt} }

\newpage
\bigskip
\hskip 3.7in\vbox{\baselineskip12pt
\hbox{NSF-ITP-00-95}
\hbox{hep-th/0008076}}

\bigskip\bigskip

\centerline{\large \bf  Gauge Dual and Noncommutative Extension}
 \centerline{\large \bf of an $\N = 2$ Supergravity Solution}

\bigskip\bigskip

\centerline{{\bf
Alex Buchel\footnote{buchel@itp.ucsb.edu},
Amanda W. Peet\footnote{peet@physics.utoronto.ca},
Joseph Polchinski\footnote{joep@itp.ucsb.edu}}}

\bigskip
\centerline{$^{1,2,3}$Institute for Theoretical Physics}
\centerline{University of California}
\centerline{Santa Barbara, CA\ \ 93106-4030, U.S.A.}

\bigskip
\centerline{$^2$Department of Physics}
\centerline{University of Toronto}
\centerline{60 St George St}
\centerline{Toronto ON  M5S 1A7 Canada}
\bigskip

\begin{abstract}
\baselineskip=16pt

We investigate some properties of a recent supergravity solution of
Pilch and Warner, which is dual to the $\N=4$ gauge theory softly
broken to $\N=2$.  We verify that a D3-brane probe has the expected
moduli space and its effective action can be brought to $\N=2$ form.
The kinetic term for the probe vanishes on an enhan\c con locus, as in
earlier work on large-$N$ $\N=2$ theories, though for the Pilch-Warner
solution this locus is a line rather than a ring.  On the gauge theory
side we find that the probe metric can be obtained from a perturbative
one-loop calculation; this principle may be useful in obtaining the
supergravity dual at more general points in the $\N=2$ gauge theory
moduli space.  We then turn on a $B$-field, following earlier work on
the $\N=4$ theory, to obtain the supergravity dual to the
noncommutative $\N = 2$ theory.

\end{abstract}
\newpage
\setcounter{footnote}{0}

\baselineskip=17pt

\sect{Introduction}\label{section:intro}

It is an important direction  to extend the AdS/CFT duality of
Maldacena~\cite{juan} to nonconformal systems with less supersymmetry.
One way to do this is by perturbing the Hamiltonian, which is
equivalent to perturbing the boundary conditions on the AdS
space~\cite{edholo,GPK}.

The understanding of the resulting solutions is still limited.  One
approach, beginning with refs.~\cite{DZ,GPPZ1}, is to reduce to
five-dimensional gauged supergravity.  This has been very useful, but
it has limitations.  For one, only rather special states can be
obtained in this way.  The solutions have only a finite number of
integration constants, whereas a gauge theory moduli space has of
order $N$ parameters.  For a second, the full ten-dimensional geometry
is in general quite complicated in the reduced directions.  This is
encoded in the five-dimensional geometry through the algebraic magic
of consistent truncation, but it is necessary to lift the solution to
ten dimensions to see its full structure.  A related issue is that
most solutions are singular.  While there have been attempts to
identify allowed singularities in a purely five-dimensional
picture~\cite{naked}, the ten-dimensional structure is crucial for a
full understanding.

For $\N = 4$ broken to $\N = 1$ or $\N = 0$ by mass terms the full
ten-dimensional geometries have recently been found~\cite{PS}.  Here
there is the simplifying feature that the 3-brane charge dominates the
dynamics, so the solution can be treated as a perturbation of the
Coulomb branch (black 3-brane~\cite{black}) solution.  However, this
approximation was found to break down in some interesting regimes.  In
particular, it becomes less useful for phases with many 5-branes.

For $\N = 4$ broken to $\N = 2$ by mass terms (the $\N = 2^*$ theory)
there is a moduli space.  Pilch and Warner (PW)~\cite{pw0004} have
recently found the ten-dimensional supergravity solution on a
one-parameter subspace of the moduli space. It is the purpose of this
paper to analyze some of the physics of the PW solution.  The PW
theory has the same {\it massless} content, pure $\N = 2$ gauge
theory, as for D7-branes wrapped on K3; the latter was studied in
ref.~\cite{jpp9911}.  In that case the naive supergravity solution had
a naked singularity, which was resolved by an interesting stringy
phenomenon.  The constituent D7 branes were forced to lie on a ring of
finite radius, the {\it enhan\c con}.  This mechanism involved states
becoming massless when the K3 on which the branes were wrapped became
small, and as such it may be a much more general phenomenon.  The PW
solution has a feature resembling the enhan\c con, and we would like
to make the connection more precise.

In section~2 we study the PW supergravity background.  We first
discuss its symmetries, and also remark on a recent $\N=1$
ten-dimensional solution~\cite{pw2}.  We then study a probe in the PW
geometry.  The basic constituents of the PW solution are the
D3-branes, and so these are the natural probes to consider.  We find
that the probe potential vanishes on a two-dimensional plane in the
transverse space, which is the correct moduli space, and that the low
energy action for the probe can be put in the expected $\N = 2$ form
by an appropriate choice of coordinates; these are checks on the PW
solution.  In addition, we will determine the precise configuration of
branes that the solution of PW represents.  We find that it is a
different part of moduli space than that studied in
ref.~\cite{jpp9911} --- the branes lie on a line segment rather than
in a ring.

In section~3 we discuss the gauge theory side of the
correspondence. We identify the $\N=2^*$ gauge theory vacuum
corresponding to linear enhan\c con of the PW geometry and compute from
the field theory perspective the moduli space metric of a D3 probe.
The $\N=2^*$ supersymmetric gauge theory was solved by Donagi and
Witten \cite{dw9510}. As we argue below, matching the supergravity
probe computation is essentially perturbative in the gauge theory, so
we will not really use the nonperturbative tools of Seiberg-Witten
theory.  The gravity and the gauge theory computations of the moduli
space metric agree up to $1/N$ corrections. This provides another
check on the proposed correspondence.

Because the gauge theory calculation is perturbative, it can be
extended to any point on the moduli space.  Thus the gauge side gives
some information about the general supergravity solution.  There is a
further simplifying feature that the gauge theory is close to the $\N
= 4$ theory, in the sense that the masses from gauge symmetry breaking
are large compared to the masses from explicit $\N = 4$ breaking; this
may allow more of the supergravity solution to be extracted.  By using
information from the gauge theory side it may be possible to find the
supergravity solution at all points on moduli space.

In section~4 we find the noncommutative generalization of the
Pilch-Warner solution, extending to PW solution the construction of
refs.~\cite{hi9907,mr9908}.  Although the PW solution is much more
complicated than $AdS_5 \times S^5$, the same strategy can be used to
generate the solution.  That is, take the $T$-dual on a $T^2$, turn on
a constant $B$-field on the $T^2$, and $T$-dualize back.  The
resulting solution should be dual to the noncommutative $\N = 2^*$
gauge theory.  As a check, we find that the D3-probe moduli space is
unaffected by the noncommutativity, a result which is expected from
the gauge theory side.

\sect{The supergravity side}\label{section:sugra}

In this section will examine the physics of the PW background by using
a D3-brane probe in  order to elucidate its properties.  The PW
background is complicated: all the IIB supergravity fields are
nontrivial.  An essentially new feature of their solution is that the
ten-dimensional dilaton-axion field depends on the radial coordinate,
and it also depends (perhaps surprisingly) on two angular transverse
coordinates as well.

\subsection{The PW solution and its symmetries}

We begin by recalling the necessary pieces of the PW
solution~\cite{pw0004}.  The ten-dimensional Einstein frame metric is
\be\ba{rl}
\bs
ds_{\rm E}^2 &\! =  {\displaystyle{
{{\left(c X_1 X_2\right)^{1/4}}\over{\rho^3}} \left\{
{{k^2\rho^6}\over{c^2-1}}dx_\parallel^2 -
{{L^2}\over{\rho^6(c^2-1)^2}}dc^2  \right. }} \cr
&\quad - {\displaystyle{ \left.
  L^2 \left[ {{1}\over{c}}d\theta^2 +
  {{\sin^2\theta}\over{X_2}}d\phi^2 + \rho^6\cos^2\theta\left(
    {{1}\over{cX_2}}\sigma_3^2 +
    {{1}\over{X_1}}(\sigma_1^2+\sigma_2^2) \right) \right] \right\} }}\ ;
\ea\label{pwmet}
\ee
the five-form field strength is
\be
\tilde F_{\rm (5)} = {\cal{F}} + \star {\cal{F}} \,, \qquad {\cal{F}} =
4 dx^0{\wedge}dx^1{\wedge}dx^2{\wedge}dx^3{\wedge}dw(r,\theta)
\,; \label{f5}
\ee
(the factor of 4 results from conventions, to be explained in
section~4.1) and the dilaton-axion is
\be
\tau =  \frac{\tau_0 - \bar\tau_0 B}{1-B}\,,
\qquad
B = e^{2i\phi} \frac{\sqrt{cX_1} - \sqrt{X_2}}{\sqrt{cX_1} + 
\sqrt{X_2}}
\,\label{pwb}
\ee
where $\tau_0 = \theta_{\rm s}/2\pi + i / \gs$ is the asymptotic value.
PW's abbreviations are\footnote{Note that
this equation for $w$ corrects a typo in PW equation (4.9), and also
extends it to general $\gs$. }
\be\label{PWabbrevs}
\ba{l}
X_1 = \cos^2\theta+c\rho^6\sin^2\theta\,,\quad
X_2 = c\cos^2\theta+\rho^6\sin^2\theta\,,\quad\cr
\displaystyle{
w(r,\theta) = \frac{k^4\rho^6X_1}{4 \gs(c^2-1)^2} }\,,
\ea
\ee
and
\be\label{rhosix}
\rho^6 = c+(c^2-1)\left[\gamma+
\frac{1}{2}\ln\left({c-1\over c+1}\right)\right]\,.
\ee
The $\sigma_i$ are the differentials
$\sigma_1=\half(\cos\alpha{}d\psi+\sin\alpha\sin\psi{}d\beta)$,
$\sigma_2=\half(-\sin\alpha{}d\psi+\cos\alpha\sin\psi{}d\beta)$,
$\sigma_3=\half(d\alpha+\cos\psi{}d\beta)$.  That is, the angles
$\alpha, \beta, \psi$ parameterize a 3-sphere, which we can also
describe by an $SU(2)$ matrix $g$ where
\be
\sigma_i = {\rm tr}(g^{-1} \tau_i dg)\ .
\ee

The above set of coordinates may be unfamiliar, so for orientation
purposes we note the behavior of various coordinates and functions of
interest in the gauge theory UV where we get back the $\N=4$ symmetry.
The AdS$_5$ metric goes as $-dr^2 + e^{2r/L}dx_\parallel^2 = -(L /
\tilde r)^2 d\tilde r^2 + ( \tilde r/ L)^2 dx_\parallel^2$, where $L =
(4\pi \gs N)^{1/4} \alpha'^{1/2}$ is the radius of curvature of
AdS$_5$ and ${\tilde{r}} = L e^{r/L}$ is the usual isotropic radial
coordinate appearing in the 3-brane harmonic function.  The asymptotic
region $r\rightarrow\infty$, $\tilde r\rightarrow\infty$ matches the
metric~(\ref{pwmet}) as $c\rightarrow 1^+$ (so $\rho^6, X_1, X_2 \to
1$), with ${\tilde{r}} = Le^{r/L} = k L/\arccosh(c)$.

The solution contains two parameters.  The parameter $k$ is
proportional to the symmetry-breaking mass perturbation $m$.  One way
to see this is to note that $k$ and $x_\parallel$ appear in the
background only in the combination $k x_\parallel$, while the gauge
theory physics depends only on the combination $m x_\parallel$.  More
precisely, eq. (63) of ref.~\cite{PS} shows that deviations from AdS
become large at $\tilde r \sim mL^2$ (where $r$ in that reference is
$\tilde r$ here), while the deviation here becomes large when $c - 1 =
O(1)$ or $\tilde r \sim kL$.  Thus $k = mL$ times a constant of order
1.

The parameter $\gamma$ defines a family of distinct solutions.  In PW,
the interpretation is given that $\gamma\ll 0$ corresponds to being on
the $\N=4$ Coulomb branch while $\gamma > 0$ is unphysical.  The
solution $\gamma = 0$ appears to have an enhan\c con, and so we are
most interested in this value but we keep $\gamma$ general for now.

The $\N = 4 \to \N = 2$ gauge theory has an $SU(2) \times U(1)$
$R$-symmetry.  The symmetry breaking gives equal masses to two of the
four Weyl fermions $\lambda_i$.  The $SU(2)$ acts on the two massless
fermions, and the $U(1) = SO(2)$ mixes the two massive fermions.  This
symmetry is evident in the metric~(\ref{pwmet}) as $g \to e^{i \zeta
\tau_3} g h$ with $h \in SU(2)$.  The six scalars transform as the
combinations $\lambda_{[i} \lambda_{j]}$ (${\bf 6} = ({\bf 4} \times
{\bf 4})_{\rm antisym}$), and two of these are invariant under $SU(2)
\times U(1)$.  Thus the supergravity solution has a fixed plane where
the radius of the transverse (squashed) two-sphere goes to zero.  At
long distance this is the equator $\theta = \pi/2$, but there is a
second coordinate patch where $\rho = 0$.  This fixed plane will play
an important role.

Note that the $SU(2) \times U(1)$ does not act on the coordinate
$\phi$.  Thus it is not surprising that the dilaton~(\ref{pwb}) has a
complicated $\phi$-dependence.  Rather, what is surprising is that the
$\phi$-dependence of $B$ is so simple, and even more surprising is
that $\phi$-translation is a Killing vector of the
metric~(\ref{pwmet}).  This can be understood as follows. The
$SL(2,\bZ)$ multiplies the fermion bilinears by a complex number; if
we extend this to $SL(2,\bR)$, then there is a $U(1)$ subgroup which
multiplies the bilinear by a phase.  A combination of this $U(1)$ and
a $U(1) \subset SO(6)$ leaves the mass perturbation invariant; call
this combination $U(1)'$.  Supergravity without branes is invariant
under $SL(2,\bR)$.  The boundary conditions are invariant under
$U(1)'$, and so in fact is the full PW solution.  Since the metric
does not transform under $SL(2,\bR)$ it is invariant under
$\phi$-translation; the field $B$ transforms by a phase under $U(1)
\subset SL(2,\bR)$ and so by a phase under $\phi$-translation.

The $\N = 2$ PW solution implicitly contains D3-branes, as we will
see, but these are invariant under $SL(2,\bR)$ and so do not affect
the forgoing argument.  Thus, the $U(1)'$ is an accidental symmetry of
the gauge theory as long as we restrict attention to states and
observables that only involve the supergravity fields and D3-branes on
the supergravity side.

It is interesting to compare the more recent $\N = 1$ solution of
Pilch and Warner (PW2)~\cite{pw2}.  Here one does expect 5-branes on
the supergravity side~\cite{PS}.  Again there is a $U(1)'$ symmetry in
the supergravity and in the boundary conditions, and some puzzling
features of the $\N = 1$ solution can be understood if this is a
symmetry of the full solution.  Namely, if there is a D5-brane as in
ref~\cite{PS}, then the $U(1)'$ will carry this into a
$(\cos\phi,\sin\phi)$-brane at angle $\phi$: the solution appears to
contain a continuous distribution of such branes on an $S^2 \times
S^1$.  Now, $(\cos\phi,\sin\phi)$-branes may sound unfamiliar, but
supergravity does not know that $(p,q)$5-brane charges are quantized,
and so it admits such sources.  Thus it might seem that the PW2
solution is illegitimate in string theory.  However, it can be
obtained as a limit of the multi 5-brane phases described in
ref.~\cite{PS}.  Namely, for large enough $n$, $n(\cos\phi,\sin\phi)$
can be approximated by integers, and such 5-branes can be distributed
around the $\phi$ direction.

\subsection{Probing the solution}

We have in general for a
D$p$-brane probe
\label{actiondbiwz}
\be
\begin{array}{rl}
\bs
S_{\rm probe} = &\! S_{\rm DBI} + S_{\rm WZ} \cr
\bs
{ } = &\! {\displaystyle{
-\mu_p \int d^{p+1}y\,  e^{-\Phi} \sqrt{
-\det\left( \bP\left[G+B\right]_{ab} 
  + 2\pi \alpha^\prime F_{ab} \right) }
}}
\cr { } &\! + {\displaystyle{
\mu_p \int \bP\left[\exp(2\pi \alpha^\prime F_{\rm (2)} +B_{\rm (2)} )
      \wedge \oplus_n C_{\rm (n)} \right] }} \,.
\end{array}
\ee
where $\mu_p^{-1}=(2\pi)^p\alpha'^{(p+1)/2}$ and $\bP$ denotes
pullback to the world-volume of the bulk fields.

We take the directions parallel to the probe to be $x_\parallel$; the
transverse coordinates $x^i$ in the conventions of PW are
$(r,\theta,\phi)$ plus the three coordinates on the (squashed) sphere
generated by the SU(2) R-symmetry of the $\N=2$ gauge theory.  In the
PW background, the components of the NS-NS or R-R two-form potentials
parallel to the D3-brane probe and the world-volume field strength all
vanish.  It follows that the only remaining terms in the probe action
come from the R-R four-form potential and from a combination of the
metric and dilaton.  The supergravity metric in \cite{pw0004} is given
in Einstein frame.  Conveniently, for the case of the D3-brane this is
precisely the combination\footnote{$G$ is the string metric.}
$g_{\mu\nu}= e^{-\Phi/2}G_{\mu\nu}$ that appears in the probe action.
The Chern-Simons terms in $F_{\rm (5)}$ do not contribute to the
longitudinal components, and so we can read the longitudinal
components of $C_{\rm (4)}$ directly from eq.~(\ref{f5}) with
${\cal{F}} = dC_{\rm (4)\parallel}$.

The probe action in static gauge becomes
\be\ba{rl}
\bs
S_{\rm probe} &\! = {\displaystyle{
\mu_3 \int d^4y
\left[ - \gs^{-1} \sqrt{-\det\bP[g_{ab}]} + \bP\left[C_{{\rm (4)}}
\right]
\right] }} \,,\cr
&\! =  {\displaystyle{
\frac{\mu_3}{\gs} \int d^4y
\left[ -\sqrt{-\det g_{ab}}
 \left(1-v^iv^j\left|g_{ij}/g_{00}\right|\right)^{1/2}
+ w(r,\theta) \right] }} \,.
\ea\ee
Inserting the PW solution, we find the potential
energy density to be
\be\label{Vprobe}
V = {{\tau_3 k^4\rho^6}\over{(c^2-1)^2}} \Bigl( \sqrt{cX_1X_2} 
- X_1 \Bigr)
\,,
\ee
with $\tau_3 = \mu_3/\gs$.
Similarly, the kinetic energy density is
\be\label{Tprobe}
\ba{rl}
\bs
T = &\! {\displaystyle{ \tau_3
 {{k^2L^2}\over{2}}{{\sqrt{cX_1X_2}}\over{(c^2-1)}} \Biggl(
{{1}\over{\rho^6(c^2-1)^2}}(v^c)^2 + {{1}\over{c}}(v^\theta)^2 +
{{\sin^2\theta}\over{X_2}}(v^\phi)^2  }} \cr
&\qquad\qquad\quad
 {\displaystyle{  + \rho^6\cos^2\theta\left\{
{{1}\over{cX_2}}(v^3)^2+{{1}\over{X_1}}\Bigl[(v^1)^2+(v^2)^2\Bigr] 
\right\}
\Biggr) }} \,,
\ea
\ee
where $v^{c,\theta,\phi,1,2,3}$ are the velocities of the probe in
the each of the six transverse directions.

For comparison to the enhan\c con physics of \cite{jpp9911}, we are
interested in the moduli space, where the potential vanishes.  There
are two solutions to the condition $V=0$,
\be
{\rm (I)}\ c X_2 = X_1 \Rightarrow \cos\theta=0 \,, \quad {\rm (II)}\
\rho^6 = 0 \,.  
\ee
Let us determine the dimensionality of these pieces of
moduli space by inspecting the kinetic terms on loci I and II.

On locus I, the kinetic term is independent of $\gamma$,
\be
T_{\rm I} = \tau_3 {{k^2L^2}\over{2}} {{c}\over{(c^2-1)}} \left[
{{1}\over{(c^2-1)^2}}(v^c)^2 + (v^\phi)^2 \right] \,.
\label{kin1}
\ee
and we see that the locus I is the $(c,\phi)$ plane.  Note that the
potential term (\ref{Vprobe}) in the D3 probe Lagrangian has a
particularly simple expansion about locus I,
\be
V_I = 0 +\tau_3 {
k^4{\rho^6}\over{2(c^2-1)}}\left(\theta-{{\pi}\over{2}}\right)^2 +
\ldots \,.
\ee

Locus II does not exist for $\gamma > 0$: the function $\rho$ is
positive on the entire range $1 < c < \infty$.  For $\gamma < 0$ there
is a unique value $c_0(\gamma)$ such that $\rho(c_0) = 0$ in
eq.~(\ref{rhosix}), and this defines locus II. Locus II is then
parameterized by $(\theta,\phi)$, and the moduli space metric there is
\be
T_{\rm II}(\gamma<0) = \tau_3 {{k^2L^2}\over{2}}
{{1}\over{( c^2_{0}-1 )}} \left[ \cos^2\theta (v^\theta)^2
+  \sin^2\theta (v^\phi)^2 \right] \,.
\ee
As noted in PW, the dilaton-axion bulk field is trivial on locus II.

For $\gamma > 0$, there is only locus I, where $c \to 1^+$ is the AdS
boundary and $c \to \infty$ is a singularity.  For $\gamma < 0$, locus
I is defined by $1 < c \leq c_0$, $\theta = \pi/2$; locus II is
defined by $c = c_0$, $0 \leq \theta \leq \pi$.  These fit together to
form a plane if we identify $\theta \cong \pi - \theta$ in locus II.
In the limit $\gamma = 0$, $c_0 \to \infty$ and locus II becomes
singular: the moduli space metric vanishes while the dilaton field
blows up.

The moduli space is two-dimensional in accordance with expectation
from $\N = 2$ gauge theory.  This is the same as the fixed plane of
the $SU(2) \times U(1)$ $R$-symmetry, consistent with the fact that
this symmetry is unbroken on the moduli space.

To identify the $\N = 2$ structure in the moduli space metric we need
also the gauge field action.  Expanding the probe action in powers of
the field strength, the coefficient of the kinetic term is $e^{-\Phi}$
and that of $F \wedge F$ is $C_{\rm (0)}$, so that
\be
\tau_{\rm YM} = 
\tau_{\rm sugra} \,.
\ee
In the natural coordinates on the $\N = 2$ moduli space, the kinetic
term for the transverse scalars is the imaginary part of $\tau_{\rm
YM}$:
\be\label{naturalcoords}
T(Y)= \frac{1}{2}\mu_3 e^{-\Phi} 
v^Y v^{\bar Y}
\ee
where $Y$ is a complex coordinate encoding the two-dimensional moduli
space.

We focus now on locus I.  From eq.~(\ref{pwb}), the dilaton is
\be\label{dilsugra}
e^{-\Phi} =
{{c}\over{\gs| \cos \phi + i c \sin \phi |^2 } }\,.
\ee
In order to find the coordinate $Y$ we first identify the obvious
isotropic coordinate $r'$ in the metric~(\ref{kin1}) via $dc/(c^2 - 1)
= -dr' / r'$.  Then in terms of
\be
z = r' e^{-i\phi} = e^{-i\phi}\sqrt{(c+1)/(c-1)}
\ee 
the locus I metric becomes
\be
\tau_3 \frac{k^2 L^2}{2} \frac{c}{(c+1)^2} v^z v^{\bar
z}\,. \label{isomet}
\ee
Equating the metrics~(\ref{naturalcoords}) and~(\ref{isomet}),
$Y$ is analytic in $z$ with
\be
\left| \frac{\partial Y}{\partial z} \right|^2 =
k^2 L^2 \left| \frac{ \cos \phi + i c \sin \phi }{c+1} \right|^2
= \frac{k^2 L^2}{4} \left| 1 - \frac{1}{z^2} \right|^2\ .
\ee
Thus
\be
Y = \frac{kL}{2} (z + z^{-1})  \ .
\ee
and
\be\label{tausug}
\tau = \frac{\tau_0 z^2 - \bar\tau_0}{z^2 - 1} = \frac{i}{\gs}
\left({{Y^2}\over{Y^2-k^2 L^2}}\right)^{1/2} + \frac{\theta_{\rm
s}}{2\pi}\ .
\ee
This is holomorphic, as expected from supersymmetry; note that $B$ is
simply $z^{-2}$.

Notice that this function has a branch cut emanating from $Y=\pm kL$.
The real line segment $-kL \leq Y \leq kL$ maps to the circle $z = 1$
and thence to $c = \infty$.  Thus the branch cut is present only for
$\gamma \geq 0$, and runs along the real axis where $c = \infty$.
This form for $\tau$ is the main result that we will need in the next
section.  Note that $kL = \zeta mL^2$ for some constant $\zeta$ which
we determine explicitly in the next section.

We would now like to know what kind of brane distribution would give
rise to the function $\tau$ which we found above.  We expect that the
source D3-branes are distributed on the Coulomb branch, and we can
infer their distribution in a number of ways.  For example, the metric
components $g_\parallel$ vanish as one approaches a D3-brane
distribution, provided that the branes are not spread in too many
dimensions.  In terms of an appropriate radial coordinate $y$, the
metric behaves as $y^{(4-k)/2} dx_\parallel^2 + y^{-(4-k)/2} dy^2$. In
the metric~(\ref{pwmet}), $g_\parallel$ vanishes only near locus II,
$\rho = 0$.  Since $\rho$ vanishes linearly at $c_0$, the metric
behaves as $(c_0 - c)^{1/2} dx_\parallel^2 + (c_0 - c)^{-3/2} dc^2$.
For $y = (c_0 - c)^{1/2}$ this is of the expected form with $k = 2$,
consistent with the two-dimensionality of locus II.  Thus, when
$\gamma < 0$ the branes are spread over locus II.  PW make the
identification that $\gamma<0$ corresponds to the Coulomb branch of
the $\N=4$ theory.  This will be true for very negative $\gamma$; for
smaller values the effect of the soft breaking parametrized by $m$
will be less negligible.

In the limit $\gamma = 0$ this locus collapses to the line segment
found above.  Curiously, the metric $c^{-2}(dx_\parallel^2 + dc^2)$ is
of $k=0$ form with $y = 1/c$, which is more singular than expected for
a one-dimensional distribution.  Evidently the effect of the
perturbation on the D3-brane metric cannot be ignored in this case.
Notice that we are at a different point in moduli space than the setup
of \cite{jpp9911}, where the branes of the enhan\c con lay on a circle
in the ``natural'' coordinates.  If we start from $\gamma=0$ and turn
on a slightly negative $\gamma$, the source brane distribution will
turn from a line segment into a very squashed disk.

Finally, for $\gamma>0$ both $g_\parallel^2$ and the string metric
$G_\parallel^2$ diverge at $c = \infty$, which appears to be
unphysical~\cite{naked,pw0004}.

The linear D3-brane distribution at $\gamma = 0$ is reminiscent of the
$\N = 2^*$ limit of the $\N = 1^*$ theory, where the 5-brane collapses
into a line as one mass is taken to zero~\cite{PS}.  However, the
length of the distribution here is $O(kL) = O(m\sqrt{\gs N}\alpha')$
in the isotropic coordinate, where the limit of a D5-brane has length
$O(m N\alpha')$ and that of an NS5-brane has length $O(m{\gs
N}\alpha')$.  The latter are both larger, reflecting the fact that the
5-branes in the $\N = 1^*$ theory are large compared with the radius
at which the perturbation becomes large.  Possibly the PW solution
could be obtained as a limit of a configuration with a large number of
5-branes.

Let us make more precise the relation of the branch cut on the real
axis to the enhan\c con of ref.\ \cite{jpp9911}.  The enhan\c con is a
distribution of D-branes on a curve where the gauge kinetic term
$e^{-\Phi}$ vanishes.\footnote{In ref.~\cite{jpp9911}, the gauge
potential term was proportional to $V - V_*$, where $V$ was the volume
of K3 and $V_*$ the self-dual volume, so that $V = V_*$ defined the
enhan\c con.}  Any further contraction of the distribution would lead
to a negative kinetic term.  From eq.~(\ref{dilsugra}), the limit $c
\to \infty$, which again is a line segment in the natural coordinates,
has vanishing kinetic term, and this is where the D3-branes are
located when $\gamma = 0$.  The shape of the distribution is dependent
on where one is on moduli space.

Let us also remark on magnetic Wilson lines.  This is relevant to the
enhan\c con physics in the following way.  In the $d=2+1$ case of
\cite{jpp9911}, the $\N=2$ gauge theory setup comes from D6-branes
wrapped on a K3.  A D0-brane probe of this system feels a force, but
the more interesting aspect of its behavior is that the coefficient of
its kinetic term goes to zero at the enhan\c con, and by duality one
can see that it becomes the gauge boson of the enhanced SU(2)
symmetry.  We would like to investigate the analog of this for the PW
system.  The most direct analogy is to consider a D-string parallel to
the D3-branes; for a static configuration we need to hang a D-string
in from infinity, and so the gauge theory dual is the magnetic Wilson
line.

We wish to concentrate on that part of the D-string worldsheet
parallel to the D3-brane.  Starting with the action
(\ref{actiondbiwz}) for a general D$p$-brane probe, we need to extract
the terms which are turned on in the PW background.  Since we are
interested only in the kinetic piece of the action we can ignore all
the Chern-Simons-type terms in $S_{\rm WZ}$.  Therefore, let us
consider the DBI piece
\be\label{donea}
S_{\rm D1} = \tau_1 \int d^2y\, e^{-\Phi}\sqrt{-\det\bP(G_{ab}+B_{ab})}
\,.
\ee
In static gauge where we allow only time dependence of the transverse
coordinates, the pullback of the NS-NS $B$-field is zero, and so we
need only the metric.  In Einstein frame, expanding to lowest order in
velocities gives
\be\label{doneb}
T_{\rm D1} = e^{-\Phi/2}\half v^i v^j \left|g_{ij}/g_{00}\right|
\sqrt{-\det(g_{ab})} = e^{-\Phi/2}\half v^i v^j g_{ij} \,,
\ee
where in the last equality we used the fact that $g_{11}=-g_{00}$ in
the PW coordinates.  The effective mass, the coefficient of the
invariant velocity, goes to zero when the dilaton blows up.  This is
the fact we were after to make the connection to enhan\c con physics.

Let us make a few remarks about the resolution of singularities by
brane expansion.  When all D3-branes are at the origin it appears that
the supergravity solutions are singular there.  When they are
sufficiently spread out then the origin is like an ordinary point and
it is possible to connect the nonnormalizable perturbation from
infinity with the normalizable solution at the origin.  An enhan\c con
distribution is one that is as compact as possible.  For the $\N =
1^*$ theory the same principle holds but there is no moduli space; the
branes are expanded by the dielectric mechanism~\cite{m9910}.

\sect{Gauge theory}

Via the AdS/CFT correspondence, the supergravity solution of
\cite{pw0004} corresponds to softly broken $\N=4$, large $N$ $\SU(N)$
Yang-Mills theory at a specific point on the Coulomb branch of the
$\N=2$ supersymmetric Yang-Mills theory with a massive adjoint
hypermultiplet. In this section we discuss the gauge theory part of
the correspondence. We identify the $\N=2$ gauge theory vacuum
corresponding to linear enhan\c con of the PW geometry at $\gamma = 0$
and compute from the field theory perspective the moduli space metric
of a D3 probe.  The supergravity calculation matches to a {one loop}
gauge theory calculation, a result that is likely to be useful in
understanding the supergravity solution at more general points in
moduli space.

In the language of four-dimensional $\N=1$ supersymmetry, the mass
deformed $\N=4$ $\SU(N)$ Yang-Mills theory consists of a vector
multiplet $V$, an adjoint chiral superfield $\Phi$ related by $\N=2$
supersymmetry to the gauge field, and two additional adjoint chiral
multiplets $Q$ and $\tQ$ which form the $\N=2$ hypermultiplet.  In
addition to the usual gauge-invariant kinetic terms for these fields,
the theory has additional interactions and hypermultiplet mass term
summarized in the superpotential\footnote{The classical K\"{a}hler
potential is normalized $(2/g_{YM}^2)\tr[\bar{\Phi}\Phi+
\bar{Q}Q+\bar{\tQ}\tQ]$.}
\be
W={2\sqrt{2}\over g_{YM}^2}\tr([Q,\tQ]\Phi)
+{m\over g_{YM}^2}(\tr Q^2+\tr\tQ^2)\,.
\label{sp}
\ee
The theory has a moduli space of Coulomb vacua parameterized by
expectation values of the adjoint scalar
\be
\Phi={\rm diag} (a_1,a_2,\cdots,a_N)\,,\quad \sum_i a_i=0\,,
\label{adsc}
\ee
in the Cartan subalgebra of the gauge group.  For generic values of
the moduli $a_i$ the gauge symmetry is broken to that of the Cartan
subalgebra $\U(1)^{N-1}$, up to the permutation of individual $\U(1)$
factors. At the semi-classical level, non-generic values of the moduli
may yield a larger symmetry group. One of the fundamental results of
the Seiberg-Witten theory~\cite{SW}\footnote{See \cite{hp9912} for an
introduction and extensive list of references.} is that in the full
quantum theory, such larger residual gauge symmetry groups do not
survive quantization, so that the theory is always in the Coulomb
phase.  The entire low energy effective action $\cL$ of the $N-1$
Abelian $\U(1)$ $\N=2$ vector multiplets is completely determined in
terms of the single prepotential $\cF\equiv\cF(\tau,m;\{a_i\})$ which
depends holomorphically on the microscopic parameters (the gauge
coupling $\tau={\theta\over 2\pi}+ i{4\pi \over g_{YM}^2}$ and the
hypermultiplet mass $m$) and the Coulomb branch moduli $\{a_i\}$
\be\ba{rl}
\bs
8\pi\cL= &\! -g_{i\bar{j}}
\left(D_\mu a^iD^\mu\bar{a}^{\bar{j}}+i\bar{\psi}\bar{\sigma}^\mu 
D_\mu \psi^i\right) \cr
& +\Re\left\{\tau_{ij}\left(
{i\over 2}F^i_{\mu\nu}F^{j\mu\nu}+
{1\over 2}F^i_{\mu\nu}\tilde{F}^{j\mu\nu}-
2\bar{\lambda}^i \bar{\sigma}^{\mu} D_{\mu}\lambda^j
\right)\right\}
\label{n2lag}
\ea\ee
with
\be
\tau_{ij}={\partial^2 \cF\over \partial a^i\partial a^j}\,,
\quad g_{i\bar{j}}=\Im[\tau_{ij}]\,.
\label{gtau}
\ee
In Eq.~(\ref{n2lag}) $\psi$'s and $\lambda$'s are fermionic
superpartners of the scalars and gauge bosons respectively.  The
covariant derivative $D_{\mu}$ is taken with respect to the
Levi-Civita connection $\Gamma^i_{jk}$ of the scalar metric
$g_{i\bar{j}}$
\be
D_\mu a^i=\partial_\mu a^i+\Gamma^i_{jk} a^j\partial_\mu a^k\,.
\ee
Classically, the prepotential is given by
\be
\cF_{\rm class}={1\over 2}\tau\sum_i a_i^2\,.
\label{fclass}
\ee
The full quantum prepotential receives both perturbative and
nonperturbative corrections
\be
\cF=\cF_{\rm class}+\cF_{\rm pert}+\cF_{\rm non-pert}\,.
\label{ffull}
\ee
The perturbative contribution is one-loop exact \cite{pertprep} and is
determined by the standard quantum field theory computation
\be\ba{rl}
\cF_{\rm pert}= & {\displaystyle{
 {i\over 8\pi}\left[\sum_{i\ne j}
\left(a_i-a_j\right)^2
\ln {\left(a_i-a_j\right)^2\over \mu^2} \right. }} \cr
& \qquad{\displaystyle{ \left. -
\sum_{i\ne j} \left(a_i-a_j+m\right)^2
\ln {\left(a_i-a_j+m\right)^2\over \mu^2}
\right] }}\,.
\label{fpert}
\ea\ee
{}From the Wilsonian effective action viewpoint it is generated by
integrating out electrically charged gauge bosons and the charged
components of the adjoint hypermultiplet.  Finally, the
nonperturbative prepotential is generated by instantons.  The
nonperturbative part of the prepotential can, in principle, be
extracted from the exact solution of the theory \cite{dw9510}. In
practice the computation is very difficult to carry out explicitly
other than for gauge groups of small rank. Nonperturbative corrections
become important in regions of moduli space with light BPS
states. Semi-classically, monopoles are expected to have masses $4\pi
v/g_{YM}^2$, where $v$ is a characteristic scale of the Higgs field
$v\sim |a_i-a_j|$.  In the $N\to\infty$ limit we scale the gauge
coupling $g_{YM}^2\to 0$ while keeping the 't Hooft coupling fixed $N
g_{YM}^2\to O(1)$.  So unless the spacing between eigenvalues of
$\Phi$ is $O(1/N)$ or smaller, instanton corrections do not survive in
this limit.

The moduli space of a D3-brane probing the PW supergravity
background is dual to the projection of the Coulomb
branch vacua of $\SU(N+1)\to \U(1)\times\U(1)^{N-1}$ to that of the
probe $\U(1)$. If $u$ is the modulus of the $\U(1)$ representing the
probe, the perturbative parametrization of the full moduli space
(\ref{adsc}) is given by
\be
\Phi={\rm diag} (u,a_1-u/N,a_2-u/N,\cdots,a_N-u/N)\,,\quad
\sum_i a_i=0 \,.
\label{adscu}
\ee
If $|u-a_i|\gg 1/N$, the instanton corrections to the metric on the
probe moduli space are exponentially suppressed and the complete
answer (in the large $N$ limit) is determined by the perturbative
prepotential. {}From (\ref{fclass}) and (\ref{fpert}) we find
\be
\tau(u)= {i\over\gs} + \frac{\theta_{\rm s}}{2\pi} + 
{i\over 2\pi }\sum_i \ln{(u-a_i-u/N)^2\over
(u-a_i-u/N)^2-m^2 }\,.
\label{mgen}
\ee
We would like to match (\ref{mgen}) and the metric on the moduli space
of the D3 probe (\ref{tausug}) in the large $N$ limit for a specific
Coulomb vacuum $\{a_1,a_2,\cdots, a_n\}$ of the ``$\U(1)^{N-1}$
background''.

Recall that the D3 probe computation in the previous section suggests
that $\N=2$ supergravity flow with $\gamma=0$ corresponds to the
Coulomb branch vacuum in which the background branes form a $\bZ_2$
symmetric linear enhan\c con singularity around the origin of the probe
moduli space.  The size of the enhan\c con in the variable $a = Y/2\pi
\alpha'$ is
\be
\frac{kL}{2\pi\alpha'} = \zeta \frac{ m L^2 }{2\pi\alpha'} = \zeta 
\frac{m \sqrt{\gs N}}{\sqrt{\pi}} \equiv a_0 \,.  
\ee
In particular, the characteristic scale of moduli space is
large compared to $m$ and so we can approximate
\be
\tau(u)= {i\over\gs} + \frac{\theta_{\rm s}}{2\pi} + 
{i\over 2\pi }\sum_i {m^2\over
(u-a_i)^2 }\,.\label{gentau}
\ee
Away
from the enhan\c con singularity nonperturbative corrections are
suppressed and the probe metric is given by the continuous limit of
this
\be
\tau(u)= {i\over\gs} + \frac{\theta_{\rm s}}{2\pi} + {i\over 2\pi}
\int_{-a_0}^{a_0}
{d a}\,
\rho(a)\frac{m^2}{(u-a)^2}
\label{metriccont}
\ee
where $\rho(a)$ is a linear density of the background
branes eigenvalues normalized as
\be
\int_{-a_0}^{a_0} da\,\rho(a)=N \,.
\label{norm}
\ee
This is to be equal to the supergravity result~(\ref{tausug}), 
\be
\tau(u) = \frac{i}{\gs}
\left({{u^2}\over{u^2- a_0^2}}\right)^{1/2} + \frac{\theta_{\rm
s}}{2\pi}\, .
\ee
Equating the discontinuities across the enhan\c con branch cut gives
\be
m^2 \rho'(u) = -\frac{2}{\gs} \frac{u}{\sqrt{ a_0^2 - u^2 }}\ ,\quad
\rho(\pm a_0) = 0\ .
\ee
This integrates to
\be
\rho(u) = \frac{2}{m^2 \gs}\sqrt{a_0^2 - u^2}\,,
\ee
and the normalization condition fixes $a_0^2 = m^2 \gs N/\pi$ or
$\zeta = 1$.

The one loop metric~(\ref{gentau}) should apply everywhere on moduli
space --- except of course when it goes negative in some region.
Seiberg and Witten~\cite{SW} showed that instanton corrections make
the metric positive everywhere.  The lesson of ref.~\cite{jpp9911} is
that at large $N$ these corrections turn on sharply on the boundary
where the metric changes sign, that is, the enhan\c con.  Outside the
enhan\c con the metric is perturbative.  The effect of nonperturbative
corrections is that the constituent D3-branes are expanded from their
perturbative positions and dissolved in the enhan\c con.

Thus the gauge theory one loop calculation gives information about the
supergravity solution anywhere on the $\SU(N)$ moduli space.
Essentially, it determines the dilaton and metric on a two-dimensional
plane.  Finding the solution on the full six-dimensional transverse
space may then be possible, with some ingenuity.  It may also be
possible to extract information about the full solution from the gauge
theory.  Off the moduli space plane supersymmetry is broken and so the
gauge theory less constrained.  However, there is a simplification in
the problem, which we have used in deriving eq.~(\ref{gentau}).  That
is, the splitting of $\N = 4$ multiplets by the mass term is small
compared to the masses from gauge symmetry breaking.  This should
restrict the renormalization of the perturbative effective action even
off the plane where supersymmetry is unbroken.

One physical interest in studying exactly solvable $\N=2$ gauge
theories is that upon deformation to $\N=1$ one hopes to get a new
handle in the mystery of confinement.  We have noted in section~2 that
the PW solution is not the $\N = 2$ limit of the confining vacuum of
ref.~\cite{PS}.  It would be very interesting to find the exact
supergravity flows corresponding to the linearized solutions of that
paper.  Constructing first the relevant $\N=2$ solution of the mass
deformed $\N=4$ Yang-Mills theory \cite{dw9510} might be a way of
approaching this problem.


\sect{Turning on a constant B-field}

Recently there has been a revival of interest in quantum field
theories formulated on noncommutative spaces, in particular those that
emerge as various limits of M-theory compactifications.  Gauge
theories are especially interesting: the limit of large
noncommutativity is similar to the large $N$ limit of ordinary gauge
theories.  In the previous section we reconstructed the low-energy
effective action on the one complex dimensional submanifold of the
moduli space of the mass deformed $\N=4$ gauge theory from its
supergravity dual. In this section we construct the deformation of the
PW flow by turning on a $B$-field on the world-volume of the D3
branes. We propose that  this deformation is the dual gravity
description of the noncommutative $\N=2$ gauge theory with massive
adjoint hypermultiplet.

After constructing the solution we consider the same observable as in
the commutative case, namely the moduli space metric for a probe
D3-brane.  In fact this metric should be the same as in the
commutative case.  The classical supergravity description is dual to
the large-$N$ limit of the gauge theory, and planar graphs in the
noncommutative theory differ from those of the commutative theory only
by a phase factor, which is trivial for the two-derivative terms which
define the moduli space metric.
    
The section is organized as follows. After fixing conventions, we
review the gravity flow dual to the noncommutative $\N=4$ gauge theory
constructed in \cite{hi9907,mr9908}. In the third part we present the
deformation of the PW flow and study the dynamics of a D3 probe in the
deformed PW geometry.

\subsection{Type to IIB equations and conventions}

We use mostly negative conventions for the signature 
$(+-\cdots -)$ and $\epsilon^{1\cdots10}=+1$.
The type IIB equations consist of \cite{schwarz83}:

\noindent $\bullet$\quad  The Einstein equations:
\be
R_{MN}=T^{(1)}_{MN}+T^{(3)}_{MN}+T^{(5)}_{MN}
\label{tenein}
\ee
where the energy
momentum tensors of the dilaton/axion field, $B$, the three index
antisymmetric tensor field, $F_{(3)}$, and the self-dual five-index
tensor field, $F_{(5)}$, are given by
\be
T^{(1)}_{MN}= P_MP_N{}^*+P_NP_M{}^*\,,
\label{enmomP}
\ee
\be
T^{(3)}_{MN}=
       {1\over 8}(G^{PQ}{}_MG^*_{PQN}+G^{*PQ}{}_MG_{PQN}-
        {1\over 6}g_{MN} G^{PQR}G^*_{PQR})
\label{enmomG}
\ee
\be
T^{(5)}_{MN}= {1\over 6} F^{PQRS}{}_MF_{PQRSN}
\label{enmomF}
\ee

In the unitary gauge $B$ is a complex scalar
field and
\be
P_M= f^2\partial_M B\,,\qquad Q_M= f^2\,{\rm Im}\,(
B\partial_MB^*)
\label{defofPQ}
\ee
with
\be
f= {1\over (1-BB^*)^{1/2}}
\label{defoff}
\ee
while the antisymmetric tensor field $G_{(3)}$ is given by
\be
G_{(3)}= f(F_{(3)}-BF_{(3)}^*)\,.
\label{defofG}
\ee

\noindent 
$\bullet$\quad The Maxwell equations:
\be
(\nabla^P-i Q^P) G_{MNP}= P^P G^*_{MNP}-{2\over 3}\,i\,F_{MNPQR}
G^{PQR}
\label{tenmaxwell}
\ee

\noindent
$\bullet$\quad The dilaton equation:
\be
(\nabla^M -2 i Q^M) P_M= -{1\over 24} G^{PQR}G_{PQR}\,
\label{tengsq}
\ee
\noindent
$\bullet$\quad The self-dual equation:
\be
F_{(5)}= \star F_{(5)}
\label{tenself}
\ee

In addition, $F_{(3)}$ and $F_{(5)}$ satisfy Bianchi identities which
follow from the definition of those field strengths in terms of their
potentials:
\bea
F_{(3)}&=& dA_{(2)}\cr
F_{(5)}&=& dA_{(4)}-{1\over 8}\,{\rm Im}( A_{(2)}\wedge
F_{(3)}^*)\,.
\label{defpotth}
\eea

The above supergravity potentials do not transform simply under
$T$-dualities. Let $\Phi$ be a dilaton, $B_{(2)}$ NSNS two-form and
$C_{(n)}$ RR forms, as conventionally defined in D-brane physics. For
the type IIB theory one would have $C_{(n)}$ with $n=0,2,4,6,8$. This
range of $n$ is consistent with the implicit summation over RR
potentials in the D-brane world-volume action.  Recall however, that
these are not all independent, but rather apper in dual
pairs. Following \cite{PS}, we define the ``modified'' field strengths
\bea
\tF_{(1)}&=&dC_{(0)}\cr
\tF_{(3)}&=&dC_{(2)}+C_{(0)} dB_{(2)}\cr
\tF_{(5)}&=&dC_{(4)}+C_{(2)}\wedge dB_{(2)}\cr
\tF_{(7)}&=&dC_{(6)}+C_{(4)}\wedge dB_{(2)}\cr
\tF_{(9)}&=&dC_{(8)}+C_{(6)}\wedge dB_{(2)}\,.
\label{fstrengths}
\eea    
The duality constraint is then implemented as 
\be
\star\tilde{F}_{(n+1)}=(-)^{n(n-1)/2} \tilde{F}_{(9-n)}\,.
\label{hodgedual}
\ee
Comparing the 
Einstein equations (\ref{tenein}) with those of \cite{PS}
we identify
\bea
C_{(0)}+ie^{-\Phi}&=&i{1+B\over 1-B}\cr
\cr
A_{(2)}&=&C_{(2)}+i B_{(2)}\cr
\cr
A_{(4)}&=&{1\over 4}\left(C_{(4)}+{1\over 2}B_{(2)}\wedge
C_{(2)}\right)\,.
\label{conv}
\eea

Now we wish to recall the $T$-duality transformations of these 
supergravity fields.  $T$-duality acts on the
Neveu-Schwarz fields as \cite{bush}:
\bea
\tilde{G}_{ yy} &=& {1\over G_{yy}}
\qquad\qquad
\qquad\qquad\qquad\qquad
e^{2 \tilde{\phi}} =\, { e^{2 \phi} \over G_{yy}}
\nonumber\\
\tilde{G}_{ \mu \nu} &=& G_{\mu \nu}
- { G_{\mu y} G_{\nu y}
- B_{\mu y} B_{\nu y}
\over G_{ yy}}
\qquad\quad
\tilde{G}_{\mu y} ={ B_{\mu y}
\over G_{yy}}
\label{NSrule}\\
\tB_{ \mu \nu}&=&B_{ \mu \nu}
-{B_{\mu y} G_{\nu y}-G_{\mu y}
B_{\nu y}\over G_{yy}}
\qquad\quad
\tB_{\mu y} ={ G_{\mu y}
\over G_{ yy}}
\nonumber
\eea
where we defined the string metric by $(G_{\alpha\beta})_{\rm string}=
e^{\Phi/2} (g_{\alpha\beta})_{\rm Einstein}$. In Eq.~(\ref{NSrule}),
$y$ denotes the Killing coordinate with respect to which the
$T$-dualization is applied, while $\mu,\nu$ denote any coordinate
directions other than $y$. If $y$ is identified on a circle of radius
$R$, \ie $y\sim y+2\pi R$, then after $T$-duality the radius becomes
$\tilde{R}=\alpha'/R=\ls^2/R$. The string coupling is also shifted as
$\tilde{g}=g\ls/R$.

$T$-duality transforms the type IIB theory into the type IIA theory
and vice versa, through its action on the world-sheet spinors
\cite{huet,leigh}. This aspect of $T$-duality is then apparent in the
transformations of the RR fields. The odd-form potentials of the IIA
theory are traded for even-form potentials in the IIB theory and vice
versa. Using the conventions adopted above, the transformation rules
for the RR potentials are \cite{meesort,m9910}:
\bea
\tilde{C}_{(n) \mu\cdots\nu\alpha y}&=&
C_{(n-1) \mu\cdots\nu\alpha}-(n-1)
{C_{(n-1) \mu\cdots\nu| y}G_{|\alpha]y}\over G_{yy}}
\nonumber\\
\tilde{C}_{(n) \mu\cdots\nu\alpha\beta}&=&
C_{(n+1) \mu\cdots\nu\alpha\beta y}
+nC_{(n-1)[\mu\cdots\nu\alpha}B_{\beta]y}\nonumber\\
&&\qquad\qquad\quad
+n(n-1){C_{(n-1)[\mu\cdots\nu|y}B_{|\alpha|y}G_{|\beta]y}\over
G_{yy}}\,.
\label{RRrule}
\eea

\subsection{Pure $\N=4$ flow}
The type IIB supergravity background dual to noncommutative $\N=4$
supersymmetric Yang-Mills theory was constructed in \cite{hi9907},
\cite{mr9908}. We briefly review this analysis here.

Generalization of quantum commutative field theories to theories on
noncommutative spaces involves adding infinitely many higher
derivative terms, which renders the theory non-local. It is thus very
hard to provide a proof of the quantum consistency of such theory
using the familiar renormalization tools of local field theories
\cite{mrs9912}. String theory provides a way to obtain noncommutative
gauge theories by considering the decoupling limit of D$(p-2)$-branes
in type II string theories on $T^2$ with a background NSNS 2-form
field $B_{\mu\nu}$ polarized along the $T^2$ \cite{CDS,DH}. The fact
that noncommutative supersymmetric gauge theory is obtained in the
decoupling limit of string theory suggests that it should be
consistent at the quantum level. In the specific example of
\cite{CDS,DH} the noncommutative gauge theory has 16 supercharges. In
general, we would expect the quantum consistency of any gauge theory
(even with less supersymmetry as in the example below), provided it
can be realized in the limit of string theory where one decouples
gravity.

Following \cite{hi9907}, consider large number of D3-branes in weakly
coupled type IIB theory, oriented along the $x^0,x^1,x^2,x^3$
directions.  Decoupling the stringy excitations by sending $\alpha'\to
0$ results in $\N=4$ supersymmetric Yang-Mills theory on the
world-volume of the D3 branes. This theory has an $AdS_5\times S^5$
supergravity dual, describing the near horizon geometry of the
D3-branes \cite{juan}. In the string frame the metric and the dilaton
is given by\footnote{We discuss RR potentials in detail in a more
general setting in the next section.}
\bea
ds_{\rm s}^2&=&{r^2\over L^2} 
\eta_{\mu\nu} dx^\mu dx^\nu-{L^2\over r^2}
dr^2-ds_5^2\cr e^{\Phi}&=&\gs
\label{ads5}
\eea
where $L^4=4\pi \gs N {\alpha'}^2$.
Consider compactifying the $x^2,x^3$ directions 
on the square torus $T^2$. The system of 
D3-branes extending along $x^0,x^1$ and wrapping the $T^2$ 
is $T$-dual to D1-branes oriented along $x^0,x^1$ directions. 
The near horizon geometry of the string frame solution in the presence
of D1-branes and their images coming from the $T^2$ compactification
is given by the $T$-dual of (\ref{ads5}) \cite{black}
\bea
ds_{\rm s}^2&=&{r^2\over L^2} (d(x^0)^2-d(x^1)^2)-{L^2\over r^2}
(d(x^2)^2+d(x^3)^2)- {L^2\over r^2} dr^2-ds_5^2\cr
e^{\Phi}&=&\gs\ {L^2\over r^2}\,.
\label{defads5}
\eea 
Let us turn on a constant $B_{(2)}$-field polarized along $T^2$.  
According to  \cite{CDS,DH} we end up in the decoupling limit 
with the noncommutative $\N=4$ Yang-Mills theory. More specifically, 
it was argued  in \cite{DH},\cite{L9802} that in order to get a finite 
noncommutative scale one should take 
\be
B_{(2)}\to\infty\,,\quad \alpha'\to 0
\label{scaling}
\ee
while keeping $B_{(2)} \alpha'$ fixed.
The constant $B_{(2)}$-field does not act as a source
for other supergravity fields, as $dB_{(2)}=0$, 
so (\ref{defads5}) with the background NSNS two form   
\be
\delta B_{(2)}=-{\triangle^2\over \alpha'}\ dx^2\wedge dx^3
\label{deltab4}
\ee 
is still a solution. $T$-duality on the $T^2$ produces 
finally the supergravity background dual to the 
noncommutative $\N=4$ Yang-Mills theory \cite{hi9907},\cite{mr9908}:
\bea
ds_{\rm s}^2&=&{r^2\over L^2} (d(x^0)^2-d(x^1)^2)-{r^2\over L^2
h}(d(x^2)^2+d(x^3)^2) -{L^2\over r^2} dr^2-ds_5^2\cr
e^{\Phi}&=&\gs/h^{1/2}\cr
\delta B_{(2)}&=&{\triangle^2 r^4\over \alpha' L^4 h}\ dx^2\wedge dx^3
\label{n4deformed}
\eea
where 
\be
h=1+{\triangle^4 r^4\over {\alpha'}^2 L^4}\,.
\label{defh}
\ee
The solution (\ref{defh}) reduces to the $AdS_5\times S^5$ solution
for small $r$, which corresponds to the IR regime of the gauge theory.
This is consistent with the field-theoretical expectations
\cite{mst0002}: the commutative $\N=4$ gauge theory does not have UV
divergences, so its noncommutative deformation does not change the IR
physics (in any case UV/IR mixing would show up only in nonplanar
effects).

\subsection{Deformed PW flow}
In constructing the gravity dual of the noncommutative $\N=2$
gauge theory we follow the  strategy of \cite{hi9907},  
reviewed above. The starting point is the PW supergravity solution 
compactified on a square torus $T^2$ along $x^2,x^3$ directions. 
The bosonic background can be written schematically as
\bea
ds_{\rm E}^2 &=& 
g_1^2\eta_{\mu\nu}dx^{\mu}dx^{\nu}-g_2^2\left(d{(x^2)}^2+d{(x^3)}^2
\right)-ds_6^2\cr
A_{(2)}&=&c_2+i b_2\cr 
F_{(5)}&=&d\chi_4+\star d\chi_4\,,\quad \chi_4= {w}\ dx^0\wedge
dx^1\wedge dx^2\wedge dx^3
\label{pworiginal}
\eea 
where $ds_6^2$ is the transverse metric and
$A_{(2)}$ has nonvanishing components only along $S^5$. The various
functions and forms depend on the six transverse coordinates and are
independent of
$x^0,\cdots, x^3$. In this initial solution the functions
$g_1$ and $g_2$ are equal. Note that the metric is given in 
Einstein frame.

$T$-duality transformations are most conveniently expressed in
fields conventional in D-brane physics. Using (\ref{conv})
we find
\bea
e^{\Phi}&=&{(1-B)(1-\bar{B})\over 1-B \bar{B}}\,,\quad C_{(0)}\equiv 
c=i{B-\bar{B}\over 
(1-B)(1-\bar{B})}\cr
\cr
B_{(2)}&=&b_2\,,\quad C_{(2)}=c_2\cr
\cr
\tF_{(5)}&=&4 F_{(5)}\,.
\label{pwdbrane}
\eea
We also define  the string frame metric according to
\bea
ds_{\rm s}^2&=&G_1^2\eta_{\mu\nu}dx^{\mu}dx^{\nu}-
G_2^2\left(d{(x^2)}^2+d{(x^3)}^2
\right)-dS_6^2\\
G_1&=& e^{\Phi/4} g_1\,,\quad G_2= e^{\Phi/4}
g_2\,,\quad dS_6^2= e^{\Phi/2} ds_6^2\,.
\label{stringmetric}
\eea 
Since 1-, 3-, and 5-form field strengths are nonzero, there will be 
nonvanishing 8-, 6-, and 4-form potentials as well
\bea
C_{(4)}&=&4w\ dx^0\wedge dx^1\wedge dx^2\wedge dx^3+\alpha_4\cr
C_{(6)}&=&f_2\wedge dx^0\wedge dx^1\wedge dx^2\wedge dx^3\cr
C_{(8)}&=&p_4\wedge dx^0\wedge dx^1\wedge dx^2\wedge dx^3
\label{864}
\eea 
where the 2-form $f_2$ and 4-forms $\alpha_4$ and $p_4$ have only
transverse components, are independent of
$x^0,\cdots,x^3$,  and satisfy
\bea
\star dc&=&(dp_4+f_2\wedge db_2)\wedge 
dx^0\wedge dx^1\wedge dx^2\wedge dx^3\cr 
\star (dc_2+c\, db_2)&=&-(d f_2+4w\, db_2)\wedge 
dx^0\wedge dx^1\wedge dx^2\wedge dx^3\cr
\star (dw \wedge dx^0\wedge dx^1\wedge dx^2\wedge dx^3)&=&
{1\over 4}\left(d\alpha_4+c_2\wedge db_2\right)\,.
\label{duality864}
\eea 
Eqs.~(\ref{duality864}) reflect the duality constraints
(\ref{hodgedual}).

Using the transformations rules (\ref{NSrule}) and (\ref{RRrule}), 
$T$-duality first along $x^3$ and then along $x^2$ 
produces the following configuration, denoted by tildes:
\bea
e^{2\tilde{\Phi}}&=&e^{2\Phi}/G_2^4\cr
\tilde{G}_1&=&G_1\,,\quad  \tilde{G}_2=1/G_2\,,\quad 
d\tilde{S}_6^2=dS_6^2\cr
\tilde{B}_{(2)}&=&B_{(2)}\cr
\tC_{(0)}&=&0\cr
\tC_{(2)}&=&c\ dx^3\wedge dx^2+4w\, dx^0\wedge dx^1\cr
\tC_{(4)}&=&c_2\wedge dx^3\wedge dx^2+f_2\wedge dx^0\wedge dx^1\cr
\tC_{(6)}&=&\alpha_4\wedge dx^3\wedge dx^2+
p_4\wedge dx^0\wedge dx^1\cr
\tC_{(8)}&=&0
\label{tildeback}
\eea
It is straightforward to verify that given (\ref{duality864}),
the field strengths constructed from the R-R potentials of 
(\ref{tildeback}) satisfy the duality constraints (\ref{hodgedual}).

As in the case of the supergravity flow corresponding to the  $\N=4$ 
Yang-Mills theory, to generate a background dual to the 
noncommutative $\N=2$ gauge theory
we now turn on  a constant NS-NS $2-$form potential on the $T^2$:
\be
\delta\tilde{B}_{(2)}=-{\triangle^2\over \alpha'}\ dx^2\wedge d x^3
\label{deltab}
\ee
Again, since the corresponding field strength vanishes, 
$\delta\tilde{B}_{(2)}$ is a modulus.  

After turning on $\tilde{B}_{(2)}$, $T$-duality along $x^2$ and then
along $x^3$ directions, followed  by the decompactification of $T^2$
produces the gravitational  dual on the noncommutative $\N=2$ gauge
theory with massive  adjoint hypermultiplet.  We denote this final
configuration with primes:
\bea
e^{2\Phi'}&=&e^{2\Phi}/h\,,\quad h=1+{\triangle^4  
G_1^4 \over {\alpha'}^2}\\[3pt]
G_1'&=&G_1\,,\quad   
G_2'=G_1/h^{1/2}\,,\quad 
 d{s_6^2}'=ds_6^2\\[3pt]
 {B_{(2)}}'&=&b_2+{\triangle^2  G_1^4\over 
\alpha' h} dx^2\wedge dx^3\cr
{C_{(0)}}'&=&c\cr
{C_{(2)}}'&=&c_2 +4{\triangle^2 w\over \alpha'}\ 
dx^0\wedge dx^1
-{\triangle^2 G_1^4\ c\over\alpha' h}\ dx^2\wedge dx^3\cr 
{C_{(4)}}'&=&{4w\over h }\
dx^0\wedge dx^1\wedge dx^2\wedge dx^3+\alpha_4+
{\triangle^2 \over \alpha'}\ f_2\wedge dx^0\wedge dx^1\cr
&&\qquad\qquad -{\triangle^2 G_1^4\over
\alpha' h}\ c_2\wedge dx^2\wedge dx^3\cr
{C_{(6)}}'&=&{1\over h}\ f_2\wedge dx^0\wedge dx^1\wedge 
dx^2\wedge dx^3+{\triangle^2 \over \alpha'}\ p_4\wedge dx^0\wedge dx^1\cr
&&\qquad\qquad -{\triangle^2  G_1^4\over
\alpha' h}\ \alpha_4\wedge dx^2\wedge dx^3\cr
{C_{(8)}}'&=&{1\over h}\ p_4\wedge dx^0\wedge dx^1\wedge 
dx^2\wedge dx^3
\label{primeback}
\eea 
where we used the fact that in the original metric 
$G_1=G_2$. We have checked that the field strengths produced 
by the RR potentials of (\ref{primeback}) satisfy the duality 
constraints (\ref{hodgedual}).

In the remaining of this section we show that a D3 probe in the 
background (\ref{primeback}) has the same moduli space 
as that in the PW geometry. Furthermore, the metric on this
moduli space is the same, as expected.

For convenience, we reproduce the action of a D3 probe 
\bea
S &=& 
-\mu_3 \int d^{4}y\,  e^{-\Phi} \sqrt{
-\det\left( \bP\left[G+B\right]_{ab} + 2\pi \alpha^\prime F_{ab} 
\right) }
\cr && + 
\mu_3 \int \bP\left[\exp(2\pi \alpha^\prime F_{\rm (2)} +B_{\rm (2)} )
      \wedge \oplus_n C_{\rm (n)} \right]  \,. 
\label{d3b}
\eea  
where $a,b$ denote directions parallel to the world volume of the
probe.  For a probe oriented along $x^0,\cdots x^3$ directions, the
potential energy density in the background (\ref{primeback}) is
\bea
V&=& \mu_3 e^{-\Phi'}
\sqrt{-\det\left(\bP[G'+B']_{ab}\right)} -\mu_3\,
\bP[{C_{(4)}}'+{C_{(2)}}'\wedge {B_{(2)}}']_{0123}\cr
&=&\mu_3\left(\gs^{-1} G_1^4 -4w \right),
\label{potb}
\eea
where to get the second line we used the transformation rules
(\ref{primeback}).  This is identical to the potential for the
original PW solution in section~2.  Thus, the moduli space of a D3
probe in the gravity background dual to the noncommutative $\N=2$
gauge theory coincides with its commutative counterpart.

Further, the metric of this space is the same.  Letting the probe
coordinates $x^i$ have a slow dependence on $y^a$, the relevant part
of the probe action is
\bea
S&=&-\mu_3\int d^4y\, e^{-\Phi'}\sqrt{-\det\left({\cal G}_{ab}
+ g_{ij}'\partial_a x^i\partial_b x^j \right) }\nonumber\\
&\stackrel{O(\partial^2)}{\to}& -\frac{\mu_3}{2}\int d^4y\,
e^{-\Phi'}\sqrt{-\det\left({\cal G}_{ab}
\right) }
\,{\cal G}^{ab}g_{ij}' \partial_a x^i\partial_b x^j 
\label{fluctuations}
\eea
where 
\be
{\cal G}_{ab} = \bP[G' + B']_{ab}\,.
\ee
Using the properties
\be
e^{-\Phi'}\sqrt{-\det\left({\cal G}_{ab}\right)}
\,{\cal G}^{(ab)} = e^{-\Phi}\sqrt{-\det\left({ G}_{ab}\right)}
\,{ G}^{ab}\,,\quad g_{ij}' = g_{ij}
\ee
of the solution~(\ref{primeback}), it follows that the metric on
moduli space is the same as in the commutative case.  We have
explained earlier why this should be true from the gauge theory point
of view.  Note that in the supergravity description this result is
obvious in the $T$-dual tilted picture, where a probe D1-brane does
not couple to the transverse $\delta B_{(2)}$.

\section*{Acknowledgements}

We wish to thank Justin David, Aki Hashimoto, Gary Horowitz, Sunny
Itzhaki, Matt Strassler, and Nick Warner for helpful discussions.
This work was supported in part by NSF grants PHY94-07194 and
PHY97-22022.


\end{document}